%% file: All_OM_QIP_final.tex
\documentclass[prl,twocolumn,showpacs,a4paper]{revtex4-1}
\usepackage{epsfig}
\usepackage{amsmath}
\usepackage{graphicx}
\usepackage{color}
\usepackage{hyperref}
\usepackage{bbm}

\newcommand{\abs}[1]{\left|#1\right|}

\newcommand{\smean}[1]{\langle #1 \rangle}

\begin{document}

\title{Optomechanical quantum information processing with photons and phonons}

\author{ K. Stannigel$^{1,2}$, P. Komar$^3$, S. J. M. Habraken$^1$,  S. D. Bennett$^3$,  M. D. Lukin$^3$, P. Zoller$^{1,2}$, P. Rabl$^1$}
\affiliation{$^1$Institute for Quantum Optics and Quantum Information, 6020
Innsbruck, Austria}
\affiliation{$^2$Institute for Theoretical Physics, University of Innsbruck, 6020 Innsbruck,
Austria}
\affiliation{$^3$Physics Department, Harvard University, Cambridge, Massachusetts 02138, USA}

\date{\today}

\begin{abstract}
We describe how strong resonant interactions in multimode optomechanical systems can
be used to induce controlled nonlinear couplings between single photons and phonons. 
Combined with linear mapping schemes between photons and 
phonons, these techniques provide a universal building 
block for various classical and quantum information processing applications. 
Our approach is especially suited for nano-optomechanical devices, 
where strong optomechanical interactions on a single 
photon level are within experimental reach.
\end{abstract}

\pacs{
42.50.Wk, 
03.67.Hk, 
07.10.Cm  
}

\maketitle
Optomechanics describes the radiation pressure interaction between an optical cavity mode and the motion of a macroscopic mechanical object, as it appears, for example, in a Fabry-P\'{e}rot cavity with a moveable mirror~\cite{OptoReview}.
First demonstrations of optomechanical (OM)  laser cooling~\cite{FirstCavityCoolingExp} have recently attracted significant interest and led
to tremendous progress in the development of new fabrication methods and experimental techniques for controlling OM interactions at the quantum level. 
Apart from ground-state cooling~\cite{TeufelNature2011,ChanNature2011}, this includes the demonstration of slow light~\cite{WeisScience2010,SafaviNaeiniNature2011}, and the coherent interconversion of optical and mechanical excitations~\cite{FiorePRL2011,VerhagenNature2012}. These achievements pave the way for a new type of quantum light-matter interface and give rise to interesting perspectives for novel OM-based quantum technologies. As a solid-state approach, such an all-OM platform would benefit directly from advanced nanofabrication and scalable integrated photonic circuit techniques. At the same time, long mechanical lifetimes comparable to those of atomic systems allow us to combine optical nonlinearities with a stationary quantum memory for light.

In this work we study strong OM coupling effects in \emph{multimode} OM systems (OMSs) and describe how resonant or near-resonant interactions in this setting allow us to exploit the intrinsic nonlinearity of radiation pressure in an optimal way. Our approach is based on the resonant exchange of photons between two optical modes mediated by a single phonon. This resonance induces much stronger nonlinearities
than achievable in single-mode OMSs, where nonlinear effects  are suppressed by a large mechanical frequency~\cite{MarshallPRL2003,LudwigNJP2008,RablPRL2011,NunnenkampPRL2011}.
Consequently, multimode OMSs provide a promising route for accessing the single-photon strong-coupling regime, where the coupling $g_0$ as well as the mechanical frequency $\omega_m$ exceeds the cavity decay rate $\kappa$~\cite{RablPRL2011}.
This regime is within reach of state-of-the-art nanoscale OM devices~\cite{ChanNature2011,EichenfieldNature2009,CarmonPRL2007,DingAPL2011} or analogous cold atom OMSs~\cite{GuptaPRL2007,BrenneckeScience2008}, and here we discuss how strong OM interactions in a multimode setup
can be used to generate single photons and 
to perform controlled gate operations 
between photonic or mechanical qubits.  
Combined with very recently developed
photon-phonon interfaces and quantum memories
based on linearized OM 
couplings~\cite{FiorePRL2011,VerhagenNature2012, Painter2011}, 
our results provide a basis for  
efficient OM classical and quantum information processing
with applications ranging from photon transistors to
quantum repeaters and networks.

\begin{figure}
\begin{center}
\includegraphics[width=0.48\textwidth]{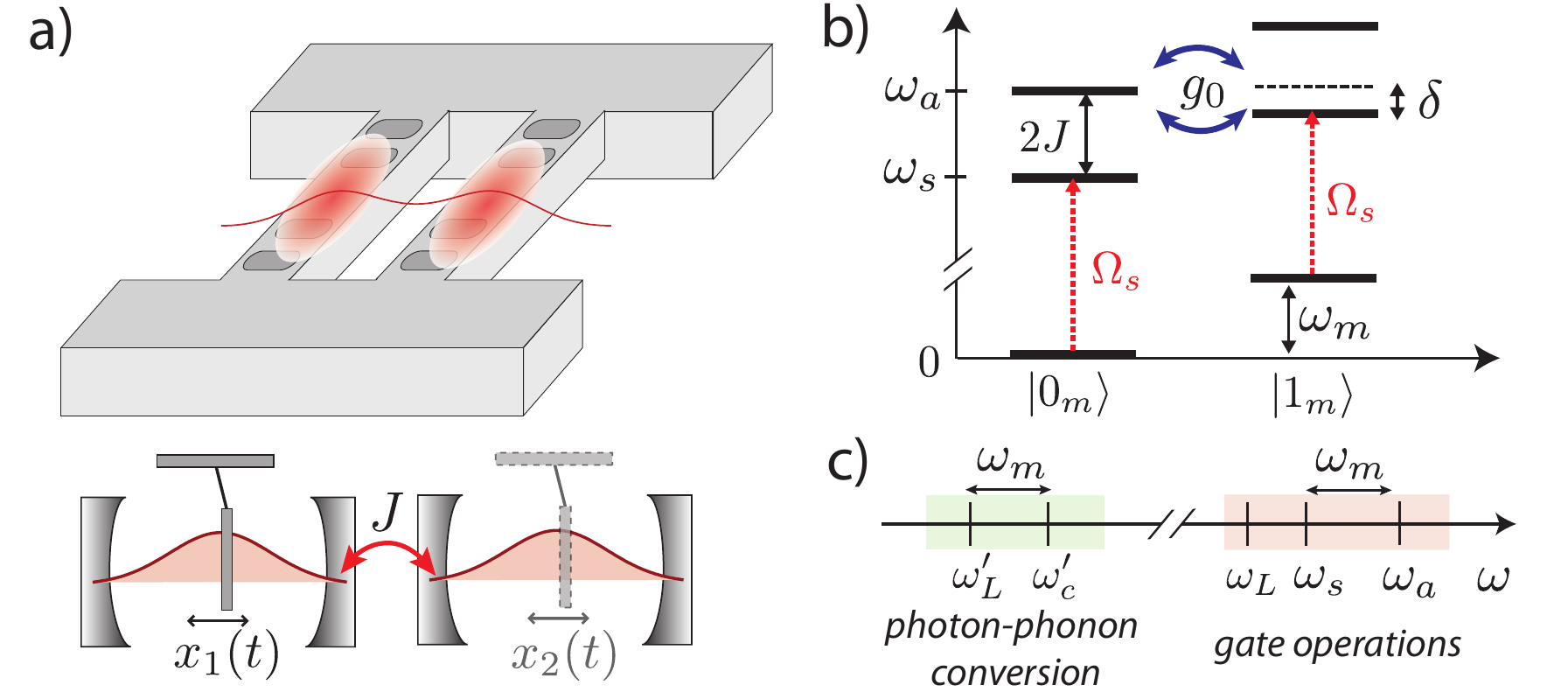}
\caption{(color online) a) Setup of two tunnel-coupled OM crystal cavities (see Ref.~\cite{EichenfieldNature2009,ChanNature2011} for more details).  b) Level diagram showing the lowest mechanical and optical excitations in a two mode OMS. Resonant coupling $(\delta=0)$ occurs when the tunnel splitting $2J$ between the optical modes is comparable to the mechanical frequency $\omega_m$. c) Different sets of strongly and weakly coupled optical modes and control laser fields can be used for nonlinear interactions $(\omega_{s},\omega_{a},\omega_L)$ and purely linear photon storage and retrieval operations $(\omega_c^\prime,\omega_L^\prime)$.  
}
\label{fig:Setup}
\end{center} 
\end{figure}

\emph{Model.} 
We consider a setup of two tunnel-coupled OMSs~\cite{MiaoPRL2009,GrudininPRL2010,DobrindtPRL2010,Painter2011,CheungPRA2011} as schematically shown in Fig.~\ref{fig:Setup}, focusing on the OM crystal design~\cite{EichenfieldNature2009,ChanNature2011} as a specific example.  
Each OMS $i=1,2$ is represented by an optical mode of frequency $\omega_c$ and a bosonic operator $c_{i}$, which is coupled via optical gradient forces to the motion of an isolated mechanical mode $b_i$ with vibrational frequency $\omega^i_m$.  The Hamiltonian for this system is $(\hbar=1)$
\begin{equation}\label{eq:H}
\begin{split}
H= &\sum_{i=1,2} \omega_m^i b_i^\dag b_i + \omega_c c_i^\dag c_i + g_0 c_i^\dag c_i (b_i+ b_i^\dag) \\
&- J (c_1^\dag c_{2} + c_{1} c_{2}^\dag) +  \sum_{i=1,2}  \Omega_i ( c_i  e^{i\omega_L t} + Ê{\rm H.c.}),
\end{split}\end{equation}
where $J$ is the tunneling amplitude between the optical modes and $g_0$ denotes the single-photon OM coupling; $\Omega_i$ are the local amplitudes of external control laser fields of frequency $\omega_L$.  Below we also consider an additional set of cavity modes and driving fields with frequencies $\omega_c^\prime $ and  $\omega_L^\prime$, respectively.  
As indicated in Fig.~\ref{fig:Setup}(c), we assume these modes to be separated in frequency and used for cooling the mechanical modes~\cite{WilsonRaePRL2007,MarquardtPRL2007}, and linear photon storage and retrieval operations~\cite{FiorePRL2011,VerhagenNature2012,ZhangPRA2003,AkramNJP2010} only.

Apart from the coherent dynamics described by Eq.~\eqref{eq:H}, we include dissipation through cavity decay and mechanical damping and model the evolution of the system density operator $\rho$ by a master equation (ME)
\begin{equation}\label{eq:ME}
\begin{split}
\dot \rho = &-i[H,\rho] + \sum_{i} \kappa \mathcal{D}[c_i] \rho + \mathcal{L}_\gamma \rho, \\
\end{split} 
\end{equation}
where  $\mathcal{D}[c]\rho=2c\rho c^\dag-\{c^\dag c, \rho\}_+$, and $\mathcal{L}_\gamma =\sum_i \frac{\gamma}{2}  (N_{\rm th}+1)  \mathcal{D}[b_i] + \frac{\gamma}{2}  N_{\rm th}  \mathcal{D}[b_i^\dag]$.
Here,  $\kappa$ is the optical field decay rate, $\gamma=\omega_m/Q$ the mechanical damping rate for a quality factor $Q$ and $N_{\rm th}=(e^{\hbar \omega_m/k_BT}-1)^{-1}$ the mechanical equilibrium occupation number for temperature $T$. Below we identify $\Gamma_m=\frac{\gamma}{2}(3N_{\rm th}+\frac{1}{2})$ as the characteristic decoherence rate for mechanical qubit states~\cite{DecoherenceRate}.

\begin{figure}
\begin{center}
\includegraphics[width=0.48\textwidth]{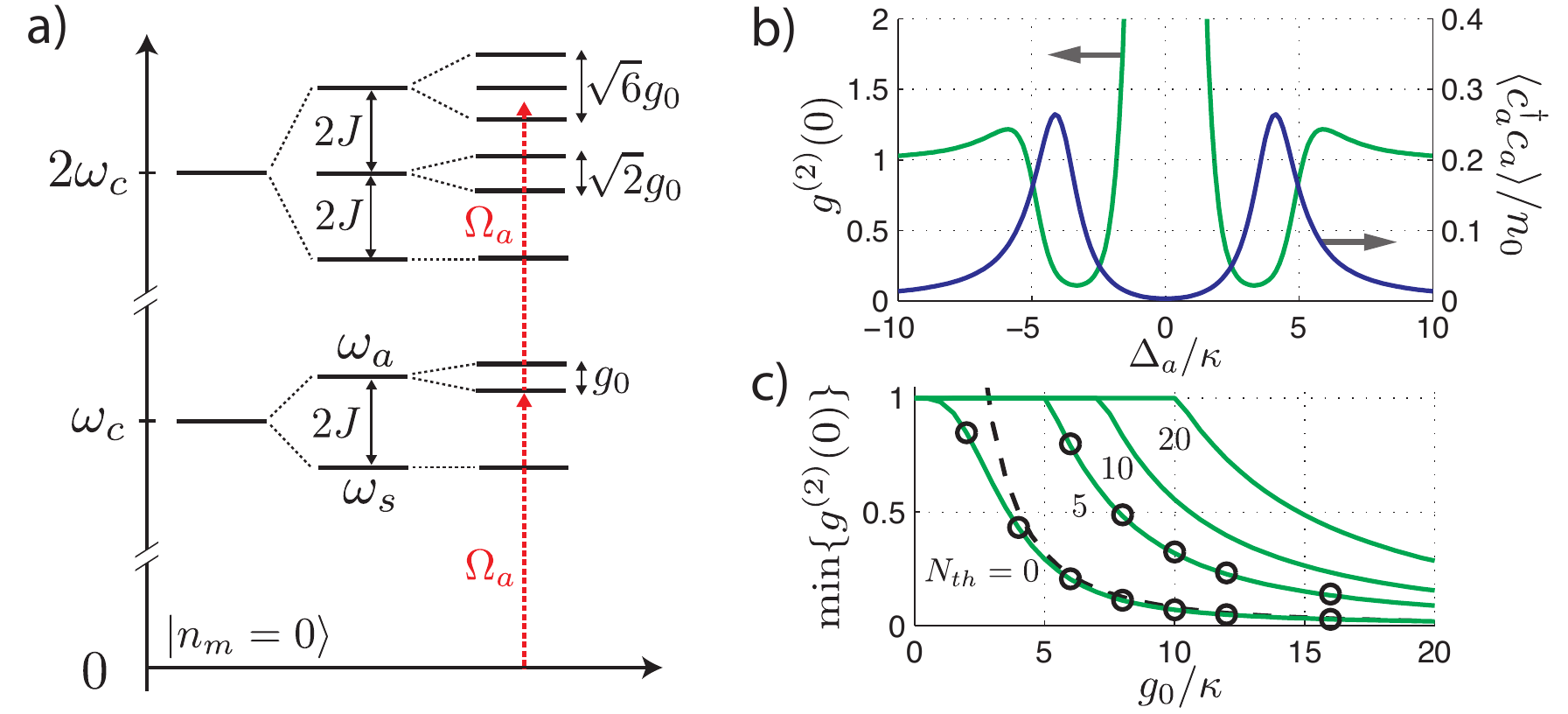}
\caption{(color online) a) Energy level diagram of a resonantly coupled OMS, $\delta=2J-\omega_m=0$, and for a single mechanical mode in the ground state. b) Excitation spectrum and $g^{(2)}(0)$ for a weak coherent field exciting the $c_a$ mode, where $g_0/\kappa=8$ and $n_0=\Omega^2_a/\kappa^2$.  c) Minimal value of $g^{(2)}(0)$  as a function of the OM coupling strength $g_0$ and for different values of $N_{\rm th}$. The analytical results (solid lines) given in the text are in good agreement with exact numerics (circles). The dashed line shows the asymptotic scaling $\sim 8\kappa^2/g_0^2$ at zero temperature.}
\label{fig:ResonantLevels}
\end{center} 
\end{figure}

\emph{Resonant strong-coupling optomechanics.} We focus on the strong coupling regime $\omega_m,g_0\gg \kappa,\Gamma_m$,  and our main goal is to show how the multimode OMS described by Eq.~\eqref{eq:H} can be used for implementing controlled interactions between qubits encoded in photonic or phononic degrees of freedom.  To illustrate this we first consider a single mechanical resonator, $b\equiv b_1$, $\omega_m\equiv \omega_m^1$. We introduce symmetric and antisymmetric optical modes $c_{s,a}=  \left( c_1 \pm c_2\right)/\sqrt{2}$ with eigenfrequencies $\omega_{s,a}$ split by $2J$. Further, we assume that $\omega_m\sim 2J \gg g_0,\kappa, |\delta|$, where $\delta=2J-\omega_m$  (see Fig.~\ref{fig:Setup}(b)). This condition can be achieved in nanoscale OMSs where $\omega_m\sim $ GHz~\cite{EichenfieldNature2009,ChanNature2011,CarmonPRL2007,DingAPL2011} and a matching tunnel splitting can be designed by appropriately adjusting the spacing between the cavities~\cite{EichenfieldNature2009,GrudininPRL2010}. In this regime we can make a rotating wave approximation with respect to the large frequency scale $\omega_m \sim 2J$ and after changing into a frame rotating with $\omega_L$ we obtain~\cite{GrudininPRL2010} 
\begin{equation}\label{eq:HRWA}
\begin{split}
H= & - \Delta_s c_s^\dag c_s - \Delta_a  c_a^\dag c_a   + \omega_m b^\dag b   \\
&+ \frac{g_0}{2} (c_a c_s^\dag b^\dag +    c_a^\dag c_s  b)+ H_\Omega(t).
\end{split}
\end{equation}
Here $ \Delta_{s,a}= \omega_L - \omega_{s,a}$ are the detunings of the driving field from the $c_s$ and $c_a$ mode, respectively, and 
$H_\Omega(t)=\sum_{\eta=s,a} \left(\Omega_\eta(t) c_\eta + {\rm H.c.}\right)$ accounts for the external driving fields with slowly varying amplitudes $\Omega_{s,a}(t)=(\Omega_1(t)\pm\Omega_2(t))/\sqrt{2}$.

The two-mode OM coupling in Eq.~\eqref{eq:HRWA} describes  photon transitions between the energetically higher mode $c_a$ to the lower mode $c_s$, while simultaneously absorbing or emitting a phonon.  For $(\Delta_s-\Delta_a-\omega_m)=\delta=0$, this leads to a resonant interaction between states $|n_a,n_s,n_m\rangle$ and $|n_a-1,n_s+1,n_m+1\rangle$, where $n_a$, $n_s$ and $n_m$ label the occupation numbers of the two optical modes and the mechanical mode, respectively.  In analogy to atomic cavity quantum electrodynamics (QED)~\cite{CavityQEDReview}, the nonlinear scaling of the corresponding transition amplitudes $\frac{g_0}{2} \sqrt{n_a (n_s+1) (n_m+1)}$ results in an anharmonic level diagram as shown in Fig.~\ref{fig:ResonantLevels}(a).  If $g_0$ exceeds the cavity linewidth $\kappa$, one and two photon transitions can be spectrally resolved, indicating the onset of strong single-photon nonlinearities.

\emph{An OM single-photon source.} 
As a potential first application of the nonlinear OM interaction we discuss the use of the OMS as a single-photon source, 
which is characterized by a vanishing equal time two-photon correlation function $g^{(2)}(0)$.  In Fig.~\ref{fig:ResonantLevels}(b) we plot the excitation spectrum $\langle c_a^\dag c_a \rangle$ and $g^{(2)}(0)=\langle c_a^\dag c_a^\dag c_a c_a\rangle/\langle c_a^\dag c_a\rangle^2$,  for the case where only the $c_a$ mode is weakly driven. Around the single-photon resonances $\Delta_a=\pm g_0/2$ we observe strong anti-bunching $g^{(2)}(0)<1$ as a clear signature of non-classical  photon statistics. To quantify this effect we assume that $\Gamma_m \ll\kappa$, which allows us to treat subspaces connected to different $|n_m\rangle$ separately.  For weak driving fields $\Omega_a\ll\kappa$, the system dynamics can then be restricted to the  six states $ |0_a,0_s, n_m\rangle, |1_a,0_s, n_m\rangle,|0_a,1_s, n_m+1\rangle,|2_a,0_s, n_m\rangle, |1_a,1_s, n_m+1\rangle,|0_a,2_s, n_m+2\rangle$ and  we calculate the relevant occupation probabilities $p_{1,0,n_m}$ and  $p_{2,0,n_m}$ to leading order in $\Omega_a$~\cite{Carmichael1991}.
We obtain
\begin{equation}
p_{1,0,n}=\abs{ \frac{4\Omega_a d}{ X_n}}^2, \qquad p_{2,0,n} = 8 \abs{ \frac{\Omega_a^2 (8 d^2 - g_0^2)}{( X_n (2X_n -g_0^2))}}^2, 
\end{equation}
where $d = \Delta_a - i\kappa$ and $X_n=d^2-g_0^2(n+1)$. By taking the appropriate thermal averages, $\smean{n_a} = \sum_n \zeta_n p_{1,0,n}$ and $g^{(2)}(0)  = 2 \sum_n \zeta_n p_{2,0,n}/\smean{n_a}^2$, where $\zeta_n=(1-e^{-\beta \hbar \omega_m})e^{-\beta \hbar \omega_m n}$ and $\beta^{-1}=k_B T$, the two photon correlation function can be evaluated for arbitrary temperatures $T$.

In Fig.~\ref{fig:ResonantLevels}(c) we plot the minimal value of $g^{(2)}(0)$ as a function of the coupling strength $g_0$ and for different $N_{\rm th}$.
As the OM coupling increases we find that for $T=0$ the minimum of the correlation function scales as $\textrm{min}_{\Delta_a}\{g^{(2)}(0)\}\simeq 8 \kappa^2/g_0^2$. This demonstrates an improved scaling over off-resonant photon blockade effects in single-mode OMSs, where for large $\omega_m$ only a small reduction $g^{(2)}(0)\simeq 1-g_0^2 /(\kappa \omega_m)$ can be obtained~\cite{RablPRL2011}. Since the positions of the single and two-photon resonances depend explicitly on the mechanical state $|n_m\rangle$,  finite temperature degrades the quality of the single-photon source.  Nevertheless, with increasing coupling strength the anti-bunching effect becomes surprisingly robust and when combined with cooling cycles to achieve $\langle n_m\rangle \sim 1$~\cite{ChanNature2011}, allows the operation of OM single-photon sources even at environmental temperatures of a few Kelvin.

\begin{figure}
\begin{center}
\includegraphics[width=0.5\textwidth]{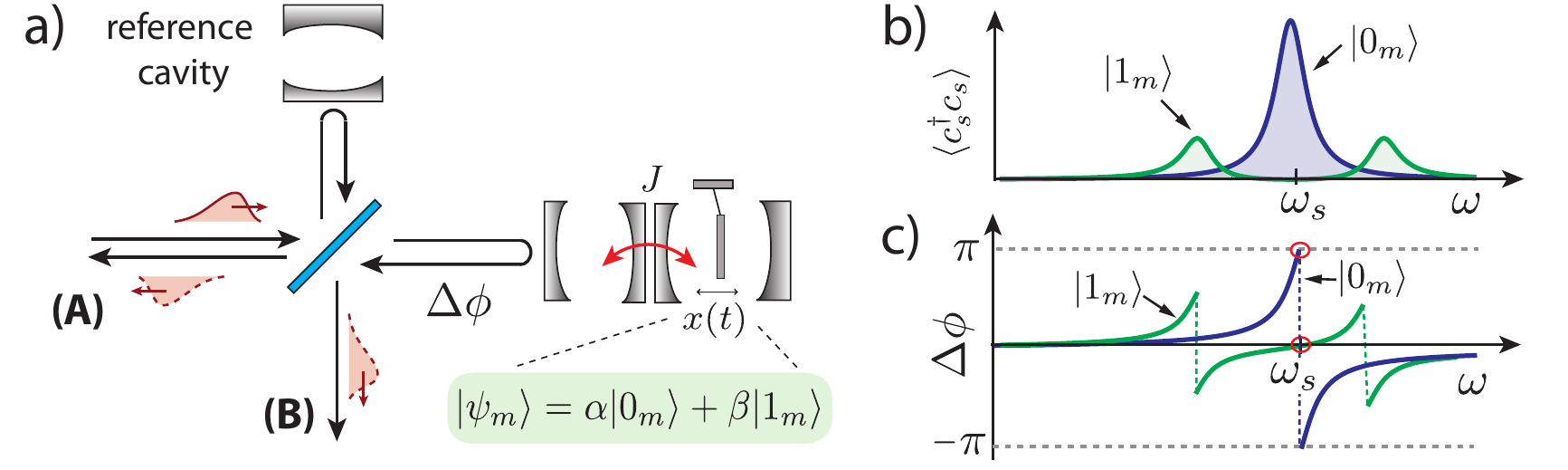}
\caption{(color online) A single-phonon single-photon transistor. a) An incoming photon in port (A) passes through the interferometric setup and leaves through port (A) or (B), depending on the phase shift $\Delta \phi$ acquired upon reflection from the two-mode OMS.   b), c) For a mechanical system in state $|0_m\rangle$, the OMS exhibits a single resonance at $\omega_s$ $(\Delta \phi=\pi$), while for state $|1_m\rangle$  the resonance splits by $g_0\gg \kappa $ and the photon does not enter the cavity $(\Delta \phi=0)$.  }
\label{fig:SinglePhononTransistor}
\end{center} 
\end{figure}

\emph{Single-phonon single-photon transistor.}  Given the ability to generate single photons, Fig.~\ref{fig:SinglePhononTransistor} illustrates a basic scheme for using the same resonant OMS to implement a two-qubit gate~\cite{DuanPRL2004}. First, we assume that the state of a control photon is mapped onto a mechanical superposition state  $\alpha|0_m\rangle+\beta |1_m\rangle$.
This can be achieved with conventional cooling 
followed by photon-phonon conversion techniques using linearized OM interactions with an auxiliary mode $\omega_c^\prime$ (see Fig.~\ref{fig:Setup}(c)). Next, a single target photon of central frequency $\sim \omega_s$ 
is sent through the interferometric setup as described in Fig.~\ref{fig:SinglePhononTransistor}. 
If the mechanical mode is in the state $|0_m\rangle$,  the incoming photon couples to a single resonant state $|0_a,1_s,0_m\rangle$ (see Fig.~\ref{fig:Setup}(b)), such that it enters the cavity and picks up a phase before being reflected.  Instead, if the mechanical resonator is in the state  $|1_m\rangle$,  the resonant coupling between $|0_a,1_s,1_m\rangle$ and $|1_a,0_s,0_m\rangle$ splits the cavity resonance, and for $g_0>\kappa$ the photon is reflected without a phase shift.
Under ideal conditions,  the final result is an entangled state
\begin{equation}\label{eq:Entangled}
|\psi\rangle = \alpha | 0_m, 1_A,0_B\rangle + \beta | 1_m, 0_A,1_B\rangle,
\end{equation}
where $A$ and $B$ are the two ports of the interferometer. This state can be converted back into an entangled state between the initial control and target photon.

Assuming that the storage and retrieval of the control photon can be achieved with high fidelity, the error for producing the entangled state~\eqref{eq:Entangled} with $\alpha=\beta=1/\sqrt{2}$ is approximately given by
\begin{equation}\label{eq:Error}
\epsilon  \approx  \frac{4\kappa^2}{g_0^2}+  \frac{1}{(\tau_p \kappa)^{2}}+  \tau_p \Gamma_m, 
\end{equation} 
where  $\tau_p$ is the duration of the single-photon pulse. The individual contributions in Eq.~\eqref{eq:Error} arise from an imperfect photon reflection, the finite spectral width of the photon pulse, and mechanical decoherence, respectively.
A minimal error is achieved for  $\tau_p^{-1}\approx \sqrt[3]{\kappa^2 \Gamma_m}$ where we obtain $\epsilon\approx {\rm  max} \{  4\kappa^2/g_0^2, \sqrt[3]{\Gamma_m^2 /\kappa^2}\}$. 
Assuming an OM crystal device with $\omega_m/(2\pi)=4$ GHz and $Q=10^5$ as discussed in Ref.~\cite{ChanNature2011}, but with an improved OM coupling $g_0/(2\pi)=50$ MHz and a lower  decay rate $\kappa/(2\pi)=5$ MHz, we obtain gate errors $\epsilon\approx 0.1$ for 
environmental temperatures around $T\approx 100$ mK.

\begin{figure}
\begin{center}
\includegraphics[width=0.5\textwidth]{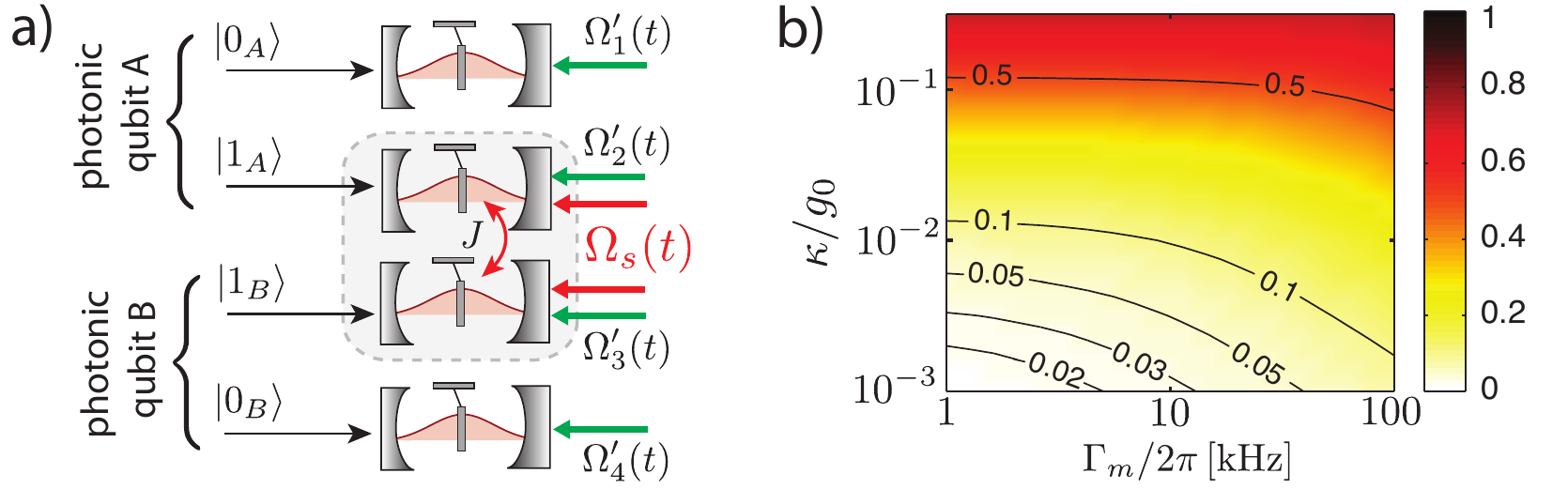}
\caption{(color online) a) OM quantum memory, where `path-encoded' photonic qubits are stored in long-lived mechanical states using tunable linearized OM interactions $\sim\Omega^\prime_i(t)$. Deterministic gate operation between stationary qubits are implemented by a controlled phonon-phonon interaction $\sim\Omega_s(t)$ as described in the text. b) The total error $\epsilon_g$ for implementing a controlled phase gate between two phononic qubits is minimized with respect to $\Delta_s$ and plotted as a function of $\kappa$ and $\Gamma_m$ (see text). The parameters for this plot are $g_0/(2\pi)=50$ MHz, $\gamma/(2\pi)=4$ kHz, $\alpha=1$ and $g_0/\delta=1/3$.    
}
\label{fig:PhononQuantumGate}
\end{center} 
\end{figure}

\emph{Phonon-phonon interactions.}  
Finally, we consider the possibility to perform a 
controlled gate operation between two qubits stored
in long-lived mechanical modes.
Our approach is depicted 
in Fig.~\ref{fig:PhononQuantumGate}(a), 
and combines the long coherence times of an OM 
quantum memory~\cite{FiorePRL2011,VerhagenNature2012,ZhangPRA2003,AkramNJP2010} 
with the practical utility of exploiting interactions 
between stationary phononic qubits.
We focus on the limit $\Gamma_m\ll \kappa$, 
and assume that optical (e.g. `path encoded') 
qubits are first mapped onto long-lived states $|0_m\rangle$ 
and $|1_m\rangle$ of two or more mechanical modes.
The OM coupling is then employed to generate 
nonlinear interactions between the phonons only.

We consider nonlinear interactions between
two mechanical modes $b_1$ and $b_2$ described
by Eq.~\eqref{eq:H}, detuned from resonance such that
$g_0< |(2J-\omega_m^i)|$ and direct transitions between  
photons and phonons are suppressed. 
To obtain the effective phonon-phonon interactions,
we first diagonalize
$H$ to second order in $\xi_i = g_0/(2J-\omega_m^i)$
with the transformation
$H\rightarrow e^{iS}He^{-iS}$, 
where 
$S=\frac{i}{2} (c_s^\dag c_a (\xi_1b_1^\dag-\xi_2b_2^\dag) -{\rm H.c.})$.
This yields
 $H= H_0 +H_g+H_\Omega(t)$, where  
 $H_0=  - \Delta_s c_s^\dag c_s - \Delta_a  c_a^\dag c_a   + \sum_i \omega^i_m b_i^\dag b_i$, 
\begin{equation}\label{eq:Hoff}
	H_g= \frac{g_0}{4} \left[ (c_s^\dag c_s\!+\!1)c^\dag_a c_a (\xi_1\!+\!\xi_2) 
	+ (c_a^\dag c_a\! -\!	c_s^\dag c_s) \mathcal{N}_b \right],
\end{equation}
and we have neglected small corrections to the driving Hamiltonian $H_\Omega(t)$.
The phonon operator in Eq.~\eqref{eq:Hoff} is given by 
$\mathcal{N}_b= \xi_1 b_1^\dag b_1+\xi_2 b_2^\dag b_2
- (\xi_1+\xi_2)(b_1^\dag b_2+ b_2^\dag b_1)/2$.
For simplicity we focus on symmetric detuning, $\omega_m^{1,2} = 2J\mp \delta$, where $\mathcal{N}_b=\frac{g_0}{\delta}(b_1^\dag b_1-b_2^\dag b_2)$. 
The  transformation also modifies the dissipative 
terms in the Eq.~\eqref{eq:ME};
most importantly, we find
an optically-induced decay channel for the mechanical modes,   
$\mathcal{L}_\gamma\rightarrow \mathcal{L}_\gamma + \kappa 
g^2_0/(4\delta^2) \mathcal{D}[c_s (b_1+b_2)]$.

We assume that only the $c_s$ mode is weakly 
driven by a slowly-varying control field $\Omega_s(t)$. 
In this case the $c_a$ mode remains unpopulated and 
we neglect it. 
Next, we shift the driven mode, $c_s\rightarrow \alpha + c_s$, 
by the classical amplitude $\alpha$,
yielding an effective ME  for $c_s$, $b_1$ and $b_2$.
Finally, we adiabatically eliminate the $c_s$ mode, valid in the
limit $|\alpha|\sim \mathcal{O}(1)$ and $(g_0^2|\alpha|/4\delta)\ll |\Delta_s+i\kappa|$,
to obtain an effective ME for the mechanical modes~\cite{SM},
\begin{equation}\label{eq:Effective}
\begin{split}
\dot \rho_m =& -i[H_m + \Lambda (b_1^\dag b_1-b_2^\dag b_2)^2 Ê, \rho_m ]  +\mathcal{L}_\gamma \rho_m\\
&+ \Gamma_\phi \mathcal{D}[(b_1^\dag b_1-b_2^\dag b_2)]\rho_m   +  \frac{\gamma^\prime}{2} \sum_i   \mathcal{D}[b_i]\rho_m.
\end{split} 
\end{equation}
Here, $\gamma^\prime=\kappa |\alpha|^2 g_0^2/(2\delta^2)$, and the phonon-phonon interaction and the phonon dephasing rate are given by
\begin{equation}
\Lambda=\frac{g_{0}^{4}|\alpha|^2  \Delta_{s}}{16 \delta^2 (\Delta_{s}^{2}+\kappa^{2})},\qquad \Gamma_\phi= \frac{g_{0}^{4}|\alpha|^2  \kappa}{16\delta^2(\Delta_{s}^{2}+\kappa^{2})}.
\end{equation}
The effective Hamiltonian in Eq.~\eqref{eq:Effective} describes 
a phonon nonlinearity with a tunable strength $\Lambda(t)\sim |\alpha(t)|^2$. 
The relevant cross-coupling is given by 
\begin{equation}\label{eq:Heff}
H_{\rm int} \simeq 2 \Lambda b_1^\dag b_1  b_2^\dag b_2,
\end{equation}
and when acting for a time $t_g=\pi/(2\Lambda)$, this 
Hamiltonian implements a controlled-phase gate between two qubits encoded in states $|0_m\rangle$ and $|1_m\rangle$. 
During this time, phonons experience intrinsic and 
optically-induced decoherence as 
seen in Eq.~\eqref{eq:Effective}.
In Fig.~\ref{fig:PhononQuantumGate}, we plot the 
resulting gate error $\epsilon_g=1-\langle\psi_0|\rho_m(t_g)|\psi_0\rangle$  
for an initial state $|\psi_0\rangle=\frac{1}{2}(|0_m\rangle+|1_m\rangle)^{\otimes2}$ 
optimized with respect to $\Delta_s$.
Using the total decoherence rate of this state, 
$\Gamma_{\rm decoh}= 2\Gamma_m+ \Gamma_\phi+ \gamma^\prime/2$, 
we find that  $\epsilon_g\propto\Gamma_{\rm decoh}/\Lambda$ 
is minimized for $|\Delta_s|\simeq g_0/2$, where $\epsilon_g\propto 4(\kappa/g_0)$. 
While this scaling with $g_0$ 
is weaker than for a gate based on photon reflection (see Eq.~\eqref{eq:Error}), 
the ability to perform a gate between stationary qubits
represents an important advantage of this approach.

\emph{Conclusions.} We have described single-photon and single-phonon nonlinear effects in strongly coupled multimode OMSs. We have shown how induced nonlinearities on or near resonance can be used for controlled quantum gate operations between flying optical or stationary phononic qubits. Our results provide a realistic route towards the quantum nonlinear regime of OMSs, and a framework for future OM information processing applications.

\emph{Acknowledgments.} The authors thank D. Chang, O. Painter and M. Aspelmeyer for valuable discussions. This work was supported by NSF, CUA, DARPA, the Packard Foundation, the EU project AQUTE and the Austrian Science Fund (FWF) through SFB FOQUS and the START grant Y 591-N16.

\emph{Note added.} During completion of this project we became aware of a related work by M. Ludwig {\it et al.}~\cite{LudwigPreprint}.

\section{Supplementary Information}

\input{ALL_OM_QIP_suppl_arxiv}

\end{document}

%% file: ALL_OM_QIP_suppl_arxiv.tex
\subsection{Phonon nonlinearities}

In Eq. (8) in the main text we have derived an effective master equation (ME) to describe the nonlinear interaction between two phonon modes. In the following we present an alternative, more rigorous, approach, which illustrates the individual approximations made in the derivation of the effective phonon nonlinearity in more detail. We first consider only a single mechanical mode, e.g. $b\equiv b_1$, which also allows us more easily to compare the results with exact numerical calculations of the full model.

\subsubsection{Model}

We start with the full ME for the two optical modes coupled to a single resonator mode, which in the frame of the driving frequency $\omega_L$ can be written as   
\begin{equation}\label{eq:SM_ME}
\dot \rho = -i[H_0+H_g+ H_\Omega(t),\rho] + \mathcal{L}_{\rm diss} \rho. \\
\end{equation}
Here 
\begin{equation}
H_0=  \omega_m b^\dag b - \Delta_s c_s^\dag c_s - \Delta_a  c_a^\dag c_a, 
\end{equation}
and
\begin{equation}
H_g= \frac{g_0}{2} \left(c_a c_s^\dag b^\dag +    c_a^\dag c_s  b \right),
\end{equation}
are the free evolution and the OM coupling, respectively,  $H_\Omega(t)= i\Omega_s(t) (c_s^\dag - c_s)$ is the driving field for the symmetric mode with slowly varying amplitude $\Omega_s(t)$ and  
\begin{equation}
\mathcal{L}_{\rm diss} \rho  = \sum_{\eta=s,a} \kappa \mathcal{D}[c_\eta] \rho + \frac{\gamma}{2} \mathcal{D}_{\rm th}[b]  \rho, 
\end{equation}
accounts for dissipation. Here we have defined the superoperator $\mathcal{D}_{\rm th}[b] = (N_{\rm th}+1)  \mathcal{D}[b] +  N_{\rm th}  \mathcal{D}[b^\dag]$ to describe the coupling to a thermal bath.

\subsubsection{Displaced frame}

In contrast to the approach outlined in the main text, we now start our analysis with a unitary displacement $U(t) c_sU^\dag (t)= c_s +\alpha(t)$ where the classical cavity field $\alpha(t)$ obeys 
\begin{equation}
\dot \alpha(t)= (i\Delta_s - \kappa) \alpha(t)  + \Omega_s(t). 
\end{equation}�
This unitary transformation eliminates the classical driving field and in the new frame the resulting ME can be written as
\begin{equation}\label{eq:SM_FullME}
\begin{split}
\dot \rho = &-i[H_{\rm lin}+H_g,\rho] + \mathcal{L}_{\rm diss} \rho, \\
\end{split} 
\end{equation}
where $H_\Omega(t)$ has disappeared, but the linear part of the Hamiltonian now contains an additional coupling between the resonator and the anti-symmetric cavity mode,
\begin{equation}\label{eq:Hlin}
H_{\rm lin}�=  H_0  + G(t) c_a b^\dag + G^*(t) c_a^\dag b,
\end{equation} 
where $G(t)=g_0 \alpha(t)/2$. Note that ME \eqref{eq:SM_FullME} is still exact and we will use this equation for our exact numerics below.

\subsubsection{Hybridized modes}

To proceed, we assume that $\alpha(t)$ is constant or slowly varying on the timescale set by the detunings $|\Delta_a+\omega_m^i|$. This allows us to write $H_{\rm lin}$ in its adiabatic eigenbasis     
\begin{equation}
H_{\rm lin}�= - \Delta_s c_s^\dag c_s  - \tilde \Delta_a C^\dag C + \tilde \omega_m B^\dag B,
\end{equation}
where  the $C$ and $B$ are bosonic operators for the hybridized mechanical and optical modes and $\tilde \Delta_a$ and $\tilde \omega_m$ are the new eigenfrequencies of $H_{\rm lin}$ for a given $G\equiv G(t)$.   We obtain
\begin{eqnarray}
C&=& \cos(\theta) c_a - \sin(\theta) b,\\
B&=& \cos(\theta) b + \sin(\theta) c_a,
\end{eqnarray}
where $\tan(2\theta)=-2|G|/\delta$ and $\delta=-(\Delta_a+\omega_m)=2J-\omega_m-\Delta_s$. The shifted frequencies are given by
\begin{eqnarray}
-\tilde \Delta_a&=& -\Delta_a - \frac{1}{2}\left( \delta - \sqrt{\delta^2+4|G|^2 }\right),\\
\tilde \omega_m &=& \omega_m - \frac{1}{2}\left( \delta +\sqrt{\delta^2+4|G|^2 }\right).
\end{eqnarray}
We see that by slowly increasing the classical control field $\alpha(t)$, the mechanical mode $b$ is adiabatically converted into a polaronic mode $B$. For small mixing angles $\theta$ the mode still retains its mechanical character, while the finite photonic component is responsible for inducing an effective nonlinearity.     

In terms of the hybridized mode operators the dissipative terms can be written as 
\begin{equation}\label{eq:SM_Ldiss}
\begin{split}
\mathcal{L}_{\rm diss}\simeq& \kappa \mathcal{D}[c_s] + \kappa \cos^2(\theta) \mathcal{D}[C] +   \frac{\gamma}{2}�\sin^2(\theta) \mathcal{D}_{\rm th}[C]  \\
+&  \frac{\gamma}{2}�\cos^2(\theta) \mathcal{D}_{\rm th}[B] + \kappa \sin^2(\theta) \mathcal{D}[B]�.
\end{split} 
\end{equation}
In particular, we identify an additional optical decay channel with rate $\gamma'=2 \kappa \sin^2(\theta)$ for the $B$ mode. In the following we define  as
\begin{equation}
\tilde{\mathcal{L}}_\gamma =   \frac{\gamma}{2}�\cos^2(\theta) \mathcal{D}_{\rm th}[B] + \frac{\gamma^\prime}{2} \mathcal{D}[B],
\end{equation}
the modified mechanical dissipation Liouvillian. 
Note that in Eq.~\eqref{eq:SM_Ldiss} we have already neglected cross-terms between $C$  and $B^\dag$. This is valid in the parameter regime considered below, where $\kappa$ is small compared to the splitting of these two modes.

Finally, we also express the nonlinear interaction $H_g$ in terms of the hybridized modes and write the result as
\begin{equation}\label{eq:SM_HgDecomp}
H_g= H_g^{(1)} + H_g^{(2)} + H_g^\prime. 
\end{equation}
Here, the first term is the one of interest 
\begin{equation}
H_g^{(1)} = \frac{g_0}{4}  \sin(2\theta)  \left( c_s + c_s^\dag\right) B^\dag B ,
\end{equation}
and describes the coupling of the $c_s$ mode to the number operator of the $B$ mode. 
The second term is given by
\begin{equation}
H_g^{(2)} = -\frac{g_0}{2}  \sin^2(\theta)  \left( B  c_s^\dag C^\dag  + B^\dag  c_s C \right),
\end{equation}
and leads to additional corrections. However, for small $\theta$ this term is small compared to $H_g^{(1)}$. It can be further reduced if  $|\Delta_s-\delta|\gg \Delta_s$.   
Finally, the last term contains interactions 
\begin{equation}
\begin{split}
H_g^{\prime} = &\frac{g_0}{2}  \cos^2(\theta)  \left( C c_s^\dag B^\dag  + C^\dag  c_s B \right) \\ & - \frac{g_0}{4}  \sin(2\theta)  \left( c_s + c_s^\dag\right) C^\dag C,
\end{split}
\end{equation}
which can be neglected when either the $c_s$ or the $C$ mode are in the vacuum state.

\subsubsection{Adiabatic elimination of the cavity mode}

Our goal is now to derive an effective ME for the mechanical degrees of freedom only. To do so, we write the full ME as 
\begin{equation}
\dot \rho = \left(\mathcal{L}_0 + \mathcal{L}_1\right)\rho,
\end{equation}
where 
\begin{equation}
\mathcal{L}_0\rho = -i[H_{\rm lin}+ H_g^\prime,\rho] + \mathcal{L}_{\rm diss} \rho, 
\end{equation}
and
\begin{equation}
\mathcal{L}_1\rho = -i[H_{g}^{(1)}+ H_g^{(2)},\rho]. 
\end{equation}
The dynamics of $\mathcal{L}_0$ does not excite the cavity modes, and therefore, in the limit where $\tilde g=g_0\sin(2\theta)/4\rightarrow 0$ (either $g_0$ is small or the mixing angle $\theta$ is small) the density operator can to a good approximation be written as $\rho(t)=\rho_m(t) \otimes \rho_c^0$, where $\rho_c^0$ is the vacuum state of the $c_s$ and the $C$ mode. To account for the effects of a small $\mathcal{L}_1\sim\tilde g$ up to second order in perturbation theory we define a projection operator onto this subspace,   
\begin{equation}
\mathcal{P}\rho = {\rm Tr}_c\{ \rho \}\otimes \rho_c^0,
\end{equation}
and its complement $\mathcal{Q}=\mathbbm{1}-\mathcal{P}$. Then
\begin{eqnarray}
\mathcal{P}\dot \rho&=& \mathcal{P}\mathcal{L}_0\mathcal{P} \rho + \mathcal{P} \mathcal{L}_{1}\mathcal{Q}\rho,\\ 
\mathcal{Q}\dot \rho&=& \mathcal{Q}(\mathcal{L}_0+\mathcal{L}_{1})\mathcal{Q} \rho +  \mathcal{Q}\mathcal{L}_{1}\mathcal{P}\rho.
\end{eqnarray}
 Up to second order in $\tilde g$ we can formally integrate the equation for $\mathcal{Q}\rho$ and obtain 
\begin{equation}
\mathcal{P}\dot{\rho}(t)\simeq\mathcal{P}\mathcal{L}_{0}\mathcal{P}\rho(t)+\mathcal{P}\mathcal{L}_{1}\int_{0}^{\infty}  d\tau\, \mathcal{Q} e^{\mathcal{L}_{0}\tau}\mathcal{Q}\mathcal{L}_{1}\mathcal{P}\rho(t).
\end{equation}
We define by $\rho_m(t)={\rm Tr}_c\{\mathcal{P} \rho(t)\}$ the reduced density operator of the mechanical mode and write the final result as
\begin{equation}\label{eq:SM_Lm}
\dot\rho_m(t)=\left( \mathcal{L}^{(0)}_m+  \mathcal{L}^{(1)}_m   +  \mathcal{L}^{(2)}_m\right) \rho_m(t).
\end{equation}
The first term describes the linear part of the dynamics  
\begin{equation}
\mathcal{L}^{(0)}_m \rho_m=  -i[ \tilde \omega_m B^\dag B,\rho_m] +  \tilde{\mathcal{L}}_\gamma \rho_m,
\end{equation} 
with a modified frequency and modified decay rates for the $B$ mode. The other two terms are given by
\begin{equation}\label{eq:SM_Lm1}
\mathcal{L}^{(1)}_m \rho_m = - \int_{0}^{\infty}  d\tau\, {\rm Tr}_c\{ [H_{g}^{(1)} , e^{\mathcal{L}_{0}\tau}\left([H_{g}^{(1)},\rho_m \otimes  \rho_c^0] \right) ]\},
\end{equation}
and
\begin{equation}\label{eq:SM_Lm2}
\mathcal{L}^{(2)}_m \rho_m = - \int_{0}^{\infty}  d\tau\, {\rm Tr}_c\{ [H_{g}^{(2)} , e^{\mathcal{L}_{0}\tau}\left([H_{g}^{(2)},\rho_m \otimes  \rho_c^0] \right) ]\}.
\end{equation}

\subsubsection{Simple perturbation theory}

In deriving Eq.~\eqref{eq:SM_Lm} we have so far only assumed that $\tilde g$ is small compared to the typical frequency scales of the dynamics of the  $c_s$ mode. For now we will also assume that $g_0$ is small compared to $\delta$ \emph{and} $\Delta_s$. This allows us to neglect the term $H_g^\prime$ in $\mathcal{L}_0$ and the cavity correlation functions in Eqs.~\eqref{eq:SM_Lm1} and \eqref{eq:SM_Lm2} can be evaluated in a straight forward manner.  For the action of $\mathcal{L}^{(1)}_m$ we obtain
\begin{equation}
\mathcal{L}^{(1)}_m \rho_m = -i  [\Lambda (B^\dag B)^2,\rho_m] + \Gamma_\phi \mathcal{D} [B^\dag B], 
\end{equation}
where $\Lambda=  {\rm Im} \{ S^{(1)}_{gg}(0) \}$,  $\Gamma_\phi=  {\rm Re}\{ S^{(1)}_{gg}(0)\}$   and 
\begin{equation}
S^{(1)}_{gg} (\omega) = \tilde g^2\int_{0}^{\infty}  d\tau\, {\rm Tr}_c\{  c_s e^{\mathcal{L}_{0}\tau} \left(c_s^\dag \rho_c^0\right)\}�e^{-i\omega \tau} .
\end{equation} 
We find $ S^{(1)}_{gg}(\omega)= \tilde g^2/(-i(\Delta_s+\omega) + \kappa)$ and after inserting back the definition of $\tilde g$ in the limit $|g_0\alpha/\delta| \ll 1$  we recover the expressions for $\Lambda$ and $\Gamma_\phi$ given in Eq. (9) in the main text.  
Similarly we obtain
\begin{equation}
\mathcal{L}^{(2)}_m \rho_m = -i  [ \delta \omega_m^{(2)}  B^\dag B,\rho_m] + \frac{\gamma^{(2)}}{2} \mathcal{D} [B], 
\end{equation}
where  $\delta \omega_m^{(2)}={\rm Im} \{ S^{(2)}_{gg}(\tilde \omega_m) \}$, $\gamma^{(2)}=  {\rm Re}\{ S^{(2)}_{gg}(\tilde \omega_m)\}$   and 
\begin{equation}
S^{(2)}_{gg} (\omega) = \frac{g_0^2 \sin^4(\theta)}{4}  \int_{0}^{\infty}  d\tau\, {\rm Tr}_c\{  c_s C e^{\mathcal{L}_{0}\tau} \left(c_s^\dag C^\dag \rho_c^0\right)\} e^{-i\omega \tau}.
\end{equation} 
The small frequency shift $\delta \omega_m^{(2)}$ can be absorbed into the definition of $\tilde \omega_m$ and, since $\gamma^{(2)}\approx \gamma^\prime \sin^2(\theta) g_0^2/(4\delta^2)$, for not too large mixing angles $\theta$, $\gamma^{(2)}$ can always be neglected compared to $\gamma^\prime$. All together the final effective phonon master equation is
\begin{equation}\label{eq:SM_EffectiveME}
\begin{split}
\dot \rho_m =& -i[\tilde \omega_m B^\dag B + \Lambda ( B^\dag B)^2 �, \rho_m ]  + \Gamma_\phi \mathcal{D}[B^\dag B]\rho_m \\
&  + \frac{\gamma}{2}\mathcal{D}_{\rm th}[B] \rho_m  +  \frac{\gamma^\prime}{2}   \mathcal{D}[B]\rho_m,
\end{split} 
\end{equation}
which is the single resonator version of ME (8) given in the main text.

\subsubsection{Corrections}

Let us now extend the above result to the case where $\tilde g$ is small compared to $\Delta_s$ and $\delta$, but the bare interaction $g_0$ is not. In this case the general expressions in Eqs.~\eqref{eq:SM_Lm1} and \eqref{eq:SM_Lm2} still apply, but the effect of $H_g^\prime$ must be taking into account when evaluating the correlation functions. To illustrate this, let us assume that $g_0$ is still small compared to $\delta$. Then, by assuming that the $C$ mode is initially in the ground state,  we obtain  approximately 
\begin{equation}
H_{\rm lin} +H_g^\prime \approx - (\Delta_s- \Delta_B B^\dag B) c_s^\dag c_s,  
\end{equation}
where  the off-resonant frequency shift is 
\begin{equation}
\Delta_B= \frac{g_0^2\cos^4(\theta)}{4(\tilde \Delta_a+\tilde\omega_m -\Delta_s))},
\end{equation}
and  can be comparable to $\Delta_s$. Therefore, we must evaluate the correlation function for each phonon number state $|n\rangle$ separately and write the resulting non-linear interaction as
\begin{equation}\label{eq:SM_EffectiveME_corr}
\mathcal{L}^{(1)}_m \rho_m = \sum_n n^2\left( -i  \left[\Lambda(n) |n\rangle\langle n|�,\rho_m\right] + \Gamma_\phi(n)  \mathcal{D} [|n\rangle\langle n| ])\right).
\end{equation}
Here $\Lambda(n)$ and $\Gamma_\phi(n)$   are the imaginary and real part of 
\begin{equation}
 S^{(1)}_{gg}(\omega=-n \Delta_B) =\frac{ n^2 \tilde g^2}{-i(\Delta_s-n\Delta_B) + \kappa}.
\end{equation}
We see that in this parameter regime more complicated nonlinearities can occur, but the overall magnitude and the ratio between coherent and dephasing interactions remains the same. In principle, this analysis can be extended to the regime, where $g_0$ is comparable to $\delta$. However, in this case no simple analytic expressions for $\lambda(n)$ and $\Gamma_\phi(n)$ can be derived and need to be evaluated numerically.

\subsubsection{Numerical simulation} 

\begin{figure*}
\includegraphics[width=0.8\textwidth]{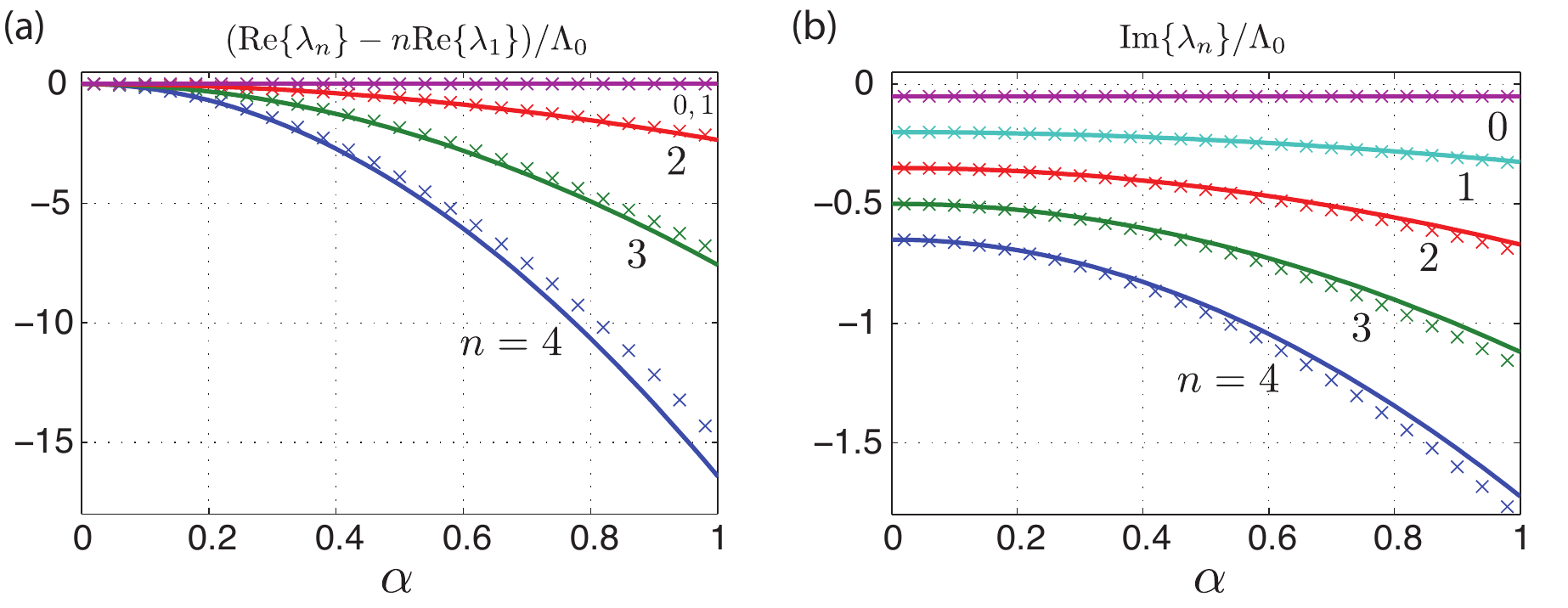}
\caption{Comparison of the effective analytic description (Eqs.\,\eqref{eq:analytics}, lines) with exact eigenvalues of the Hamiltonian in Eq.\,\eqref{eq:Hfull} (crosses) for different cavity field amplitudes $\alpha$. All results are normalized to the scale $\Lambda_0=g_0^4/(16\abs{\Delta_s}\delta^2)$ of the non-linearity.
(a) Deviation of the real parts of the eigenvalues from the result expected for a linear oscillator, such that the splitting of the curves indicates an effective non-linearity.
(b) Imaginary parts of the eigenvalues corresponding to decays. In both plots we used the parameters $\Delta_s/g_0=-1$, $\delta/g_0=5$, $\kappa/g_0=2.5\times 10^{-2}$, $\gamma_m/g_0=2.5\times 10^{-4}$ and $N_{\rm th}=1$.}
\label{fig:numerics}
\end{figure*}

To assess the validity of the effective phonon ME we now compare our result with the dynamics of the full OMS.  Since we are mainly interested in the relation between the phonon non-linearity and the corresponding dephasing and decay rates, it is sufficient to evaluate the spectrum of the non-Hermitian Hamiltonian, which  for the full model it is given by
\begin{equation}
\label{eq:Hfull}
\begin{split}
\tilde H_{\rm full}= & H_{\rm lin}+ H_g - i\kappa c_s^\dag c_s - i\kappa c_a^\dag c_a\\
& -i\frac{\gamma}{2} (N_{\rm th}+1) b^\dag b  -i\frac{\gamma}{2} N_{\rm th} b b^\dag.  
\end{split}
\end{equation}
In Fig.~\ref{fig:numerics} we plot the real and imaginary parts of the lowest eigenvalues $\lambda_n$ of $\tilde H_{\rm full}$, which correspond to the lowest number states $|n\rangle$ of the $B$ mode.
From the effective phonon model given in Eq.~\eqref{eq:SM_EffectiveME} and \eqref{eq:SM_EffectiveME_corr} we obtain the approximate analytic results
\begin{subequations}
\label{eq:analytics}
\begin{equation}
{\rm Re} \{ \lambda_n\}= n \tilde \omega_m + n^2 \Lambda(n),
\end{equation}  
and
\begin{equation}
|{\rm Im} \{ \lambda_n\}|=\frac{\gamma}{2} N_{\rm th} + n \left( \frac{\gamma}{2} (2N_{\rm th}+1) +\frac{\gamma^\prime}{2}\right) +n^2 \Gamma_\phi(n). 
\end{equation}
\end{subequations}
We see a good agreement between these results for the effective model and the exact numerics, both for the real and imaginary parts. Although there are some deviations due to higher-order effects, the effective non-linear splitting (Fig.~\ref{fig:numerics}(a)) is much larger than the induced decoherence (Fig.~\ref{fig:numerics}(b)), as is expected for the chosen parameters. Hence, we conclude that the effective model accurately describes the dynamics of the mechanical resonator, and that the effective phonon non-linearity may serve as a basis for gate operations as discussed in the main text and in the following section.

\subsection{Phonon-phonon interactions} 

The derivation of the effective phonon nonlinearity, as outlined above for a single resonator, can be easily adapted to two resonators as discussed in the main text.  
In this case we have 
\begin{equation}
H_0=  \sum_{i=1,2} \omega^i_m b_i^\dag b_i - \Delta_s c_s^\dag c_s - \Delta_a  c_a^\dag c_a, 
\end{equation}
and
\begin{equation}
H_g= \frac{g_0}{2} \left[c_a c_s^\dag (b^\dag_1-b_2^\dag) +    c_a^\dag c_s  (b_1-b_2)\right].
\end{equation}
After changing into the displaced representation to eliminate the driving field we obtain the linearized Hamiltonian   
\begin{equation}
H_{\rm lin}�=  H_0  + \sqrt{2}\left( G(t)  c_a b_a^\dag + G^*(t) c_a^\dag b_a\right),
\end{equation} 
where $b_a=(b_1-b_2)/\sqrt{2}$ and $G(t)=g_0\alpha(t)/2$. 
For similar mechanical frequencies $\omega_m^1\simeq \omega_m^2=\omega_m$ the symmetric resonator mode is decoupled and we can simply repeat the analysis from above by identifying $b\equiv b_a$ and replacing $g_0$ by $\sqrt{2}g_0$. 

For arbitrary $\omega_i$, we write the linear part of the Hamiltonian in its diagonal form 
\begin{equation}\label{eq:Hlin}
H_{\rm lin}�=  - \Delta_s c_s^\dag c_s -  \tilde \Delta_a  C^\dag C + \tilde \omega_1 B_1^\dag B_1 + \tilde \omega_2 B_2^\dag B_2.
\end{equation} 
As in the single-resonator case the $c_s$ mode is unaffected,  but the $c_a$ mode now couples to both $b_1$ and $b_2$.  The resulting hybridized modes $C$, $B_\pm$ depend on the choice of parameters $\omega_m^{1,2}, \Delta_a$ and $G$. For the case of interest, i.e. for a symmetric detuning $\omega_m^{1,2}=-\Delta_a \mp \delta$, we obtain
\begin{eqnarray}
C&=& \cos(2\Theta) c_a -  \sin(2\Theta)(B_1 + B_2)/\sqrt{2},\\
B_1&=& \cos^2(\Theta) b_1 + \sin(2\Theta) c_a/\sqrt{2} -  \sin^2(\Theta)b_2,\\
B_2&=& \cos^2(\Theta) b_2 + \sin(2\Theta) c_a/\sqrt{2} -  \sin^2(\Theta)b_1,
\end{eqnarray} 
where $\tan(2\Theta)=-\sqrt{2} |G|/\delta$. Therefore, for small $\Theta$ the modes $B_{1,2}$ correspond to the original mechanical resonator modes $b_{1,2}$ and $\tilde \omega_i\approx \omega_m^{i}$.

As above, we can now re-express the dissipation and the non-linear coupling $H_g$ in terms of $C$ and $B_\pm$. The modified mechanical dissipation terms are given   
\begin{equation}
\tilde{\mathcal{L}}_\gamma =  \sum_{i=1,2}  \frac{\gamma}{2}�\cos^2(2\Theta) \mathcal{D}_{\rm th}[B_i] + \frac{\kappa}{2} \sin^2(2\Theta) \mathcal{D}[B_i],
\end{equation}
and for small $\Theta$ the optical decay rate $\gamma^\prime =\kappa \sin^2(2\Theta)$ is the same as given above and in the main text.  
Using the decomposition of the non-linear coupling as done in Eq.~\eqref{eq:SM_HgDecomp},  we obtain 
\begin{equation}
H_g^{(1)}= \frac{g_0}{\sqrt{8} }\sin(2\Theta)   \left( c_s+c_s^\dag\right) \left(B_1^\dag B_1 - B^\dag_2 B_2\right),
\end{equation} 
the contribution $H_g^{(2)}$ vanishes and 
\begin{equation}
H_g^{\prime}= \frac{g_0}{2  }\cos(2\Theta)   \left( C c_s^\dag (B_1^\dag-B_2^\dag) + {\rm H.c.}  \right).
\end{equation}
We see that the structure and also the relative frequency scales are identical to the corresponding terms discussed for the single resonator above. Therefore, under the same conditions we can eliminate the cavity mode and  obtain the effective phonon master equation
\begin{equation}
\begin{split}
\dot \rho_m =& -i\left[\sum_i \tilde \omega_i B_i^\dag B_i + \Lambda (B_1^\dag B_1-B_2^\dag B_2)^2 �, \rho_m \right]  \\
&+ \Gamma_\phi \mathcal{D}[(B_1^\dag B_1-B_2^\dag B_2)]\rho_m   + \tilde{\mathcal{L}}_\gamma \rho_m.
\end{split} 
\end{equation}
 For small $\Theta$ this equation reduces to ME  (8) in the main text and higher-order corrections can be included in the same way as discussed for the single resonator case.